\documentclass[aps,prl,twocolumn,superscriptaddress,amsmath,showpacs]{revtex4}
\usepackage{epsfig}

\begin{document}
\preprint{\vtop{\hbox{RU04-2-B}\hbox{MCTP-02-06}\hbox{hep-th/0406031}
\vskip24pt}}

\title{Gauge Field Fluctuations and First-Order Phase Transition in Color
Superconductivity}

\author{Ioannis Giannakis}
\email{giannak@summit.rockefeller.edu}
\affiliation{Physics Department, The Rockefeller University,
1230 York Avenue, New York, NY 10021-6399}
\author{Defu Hou}
\email{hou@th.physik.uni-frankfurt.de}
\affiliation{Institut f\"ur Theoretische Physik, 
Johann Wolfgang Goethe--Universit\"at,
D-60054 Frankfurt am Main, Germany}
\affiliation{Institute of Particle Physics, Huazhong Normal University,
Wuhan, 430079, China}
\author{Hai-cang Ren}
\email{ren@summit.rockefeller.edu}
\affiliation{Physics Department, The Rockefeller University,
1230 York Avenue, New York, NY 10021-6399}
\author{Dirk H.\ Rischke}
\email{drischke@th.physik.uni-frankfurt.de}
\affiliation{Institut f\"ur Theoretische Physik, 
Johann Wolfgang Goethe--Universit\"at,
D-60054 Frankfurt am Main, Germany}

\begin{abstract}
We study the gauge field fluctuations in dense quark matter and 
determine the temperature of the induced first-order phase transition to
the color-superconducting
phase in weak coupling. We find that the local approximation of the
coupling between the gauge potential and the order parameter, employed in 
the Ginzburg-Landau theory, has to be modified by restoring the full momentum 
dependence of the polarization function of gluons in
the superconducting phase. 
\end{abstract}

\maketitle


Quantum chromodynamics (QCD) at high baryon density has become
an active research area in recent years \cite{love,reviews,review}. 
The interest has been focussed on exploring the nature of nuclear
matter under extreme conditions of temperature and density.
The physics of this region of the phase diagram of QCD is
relevant to the phenomenology of high-energy nuclear collisions and to the
properties of highly compressed nuclear matter inside a compact star.

Despite the fact that a nonperturbative approach is lacking --- lattice
simulations are notoriously difficult because of the fermion
sign problem --- a number of interesting results have been obtained 
based on Nambu--Jona-Lasinio (NJL)--type models which are
treated in the mean-field approximation \cite{NJLreview}.
At these high densities --- several times higher than
the density of ordinary nuclear matter ---
a novel, {\em color-superconducting\/} phase of nuclear matter
is expected to appear \cite{love,arw}.
 
Since QCD is an asymptotically free theory,
reliable perturbative calculations can be performed for 
asymptotically large quark chemical potentials, $\mu \gg \Lambda_{\rm QCD}$.
Thus, at asymptotic densities
color superconductivity can be quantitatively explored within QCD.
At such ultra-high chemical potentials, the interaction between
two quarks near their Fermi surface is dominated by one-gluon
exchange, which is attractive in the color-antisymmetric channel
for both the color-electric and color-magnetic parts. The formulas
for the energy gap and the transition temperature have been derived
in Refs.\ \cite{son,schafer,rischke,hong,brown,wang}. In weak coupling and
for three colors and three flavors of quarks, the critical
temperature, $T_c$ that corresponds to the pairing 
instability of the normal phase is given by
\begin{equation}
\label{pairing}
\ln\frac{k_BT_c}{\mu}=-\frac{3\pi^2}{\sqrt{2}g}
+\ln\frac{2048\sqrt{2}\pi^3}{9\sqrt{3}g^5}
+\gamma-\frac{\pi^2+4}{8}+O(g).
\end{equation}
The color superconductor in this region is of type I. 

The physics of a color superconductor in the vicinity of the
transition temperature can be described in terms of
the Ginzburg-Landau free energy functional
\cite{lida,ioannis},
which depends on the expectation values of the order
parameter and the gauge potential. As the temperature, $T$, gets
sufficiently close to the critical temperature
$T_c$, the fluctuations of the order parameter and of the gauge
potential cannot be ignored. Within the framework of the
Ginzburg-Landau approach, we can estimate 
the free energy density of the fluctuations
by the thermal energy $k_B T$ within
a volume $l^3$, where $l$ is the characteristic length of the
fluctuation. This volume would be $\xi^3$ for the order parameter
and $\lambda^3$ for the gauge potential, 
$\xi$ and $\lambda$ being the coherence length and the
magnetic penetration depth at temperature $T$, respectively. Since both
lengths diverge like $|T-T_c|^{-1/2}$, the corresponding fluctuation 
energy density behaves as $|T-T_c|^{3/2}$. The condensation energy density,
however, behaves as $(T-T_c)^2$. Therefore, as $T\to T_c$, 
the fluctuation energy density will eventually dominate and the nature of the
phase transition will be modified. For a strong type-I superconductor,
$\lambda \ll \xi$, and the fluctuations of the gauge field
exceed by far the fluctuations of the order parameter. 
It is then permissible to retain the fluctuations of the gauge field 
while neglecting those of the order parameter. As we shall see, 
a first-order phase transition occurs at $T_c^*>T_c$, 
while the temperature $T_c$, determined by 
the pairing instability, represents 
the lower bound for a supercooled normal phase without nucleation.

The strength of the first-order phase transition is measured by the
ratio $(T_c^*-T_c)/T_c$, where $T_c^*$ and $T_c$ are
the temperatures of the first- and second-order phase
transitions, respectively. A subtle issue arises 
in the calculation of the transition
temperature $T_c^*$, due to the significance of the
relation between the magnetic penetration depth
$\lambda$, which characterizes the momentum dependence of the 
self-energy of magnetic gluons in the superconducting phase, and the
coherence length {\it{at zero temperature}}, $\xi_0\sim 1/(k_B T_c)$.
The validity of the local coupling between the order parameter
and the gauge potential employed in the Ginzburg-Landau approach
relies on the inequality
\begin{equation} \label{ineq}
\xi_0 \ll \lambda \ll \xi
\end{equation} 
being valid at $T_c^*$. The original
calculation of Bailin and Love \cite{love},
which generalizes the method of Halperin, Lubensky and Ma \cite{ma} for a 
metallic superconductor, as well as the recent one of Ref.\ \cite{gia},
ignored this subtlety and overestimated $T_c^*$.
Recently, the authors of Ref.\ \cite{baym} took into account 
the momentum dependence of the gluon mass by truncating the contribution
of the fluctuation modes with momentum
higher than $\xi_0^{-1}$ and obtained a weaker first-order transition. 
Both approaches employ the local-coupling approximation, but their results
are not completely consistent with Eq.(\ref{ineq}).

In this paper
we have restored the full momentum dependence of the magnetic gluon 
self-energy in our calculation. Due to the forward singularity of 
one-gluon exchange, a determination of the strength of the first-order
phase transition can be achieved within the framework of QCD.
We find that in the weak-coupling limit
\begin{equation}
\frac{T_c^*-T_c}{T_c}=\frac{\pi^2}{12\sqrt{2}}\, g \simeq 0.58\, g\;.
\label{eqfirst0}
\end{equation}
Contrary to the assumption (\ref{ineq}) made in previous works, it then follows 
that, at $T_c^*$,
\begin{equation}
\lambda \ll \xi_0 \ll \xi\;.
\end{equation}

The effect of the gauge fluctuations
can be incorporated into the free energy of dense 
quark matter around $T_c$ by using the Cornwall-Jackiw-Tomboulis (CJT) 
formalism \cite{cjt}.  Denoting the full 
gluon and quark propagators by
${\cal D}$ and ${\cal S}$, respectively, and regarding them as
variational parameters, for a homogeneous system
we write the CJT effective potential as
\begin{eqnarray}
\lefteqn{ \Gamma[{\cal D}, {\cal S}]= \frac{k_B T}{2 \,\Omega}
\left\{ {\rm Tr}\, \ln {\cal D}^{-1} +  {\rm Tr}\,
(D^{-1}{\cal D}-1)   \right. }\nonumber \\
& - & \left.  {\rm Tr}\, \ln{\cal S}^{-1}
-  {\rm Tr}\, (S^{-1}{\cal S}-1)
- 2\,\Gamma_{2}[{\cal D}, {\cal S}] \right\}\;,
\label{eqtik}
\end{eqnarray}
where $\Omega$ is the 3-volume of the system,
$D^{-1}$ and $S^{-1}$ are the inverse tree-level
propagators for gluons and quarks, respectively, and
$\Gamma_2$ represents the 
sum of all $2$PI vacuum diagrams built with ${\cal D}$ and ${\cal S}$
\cite{review}. Thermal equilibrium corresponds to
\begin{equation}
\frac{\delta\Gamma}{\delta{\cal D}}=0\quad, \qquad
\frac{\delta\Gamma}{\delta{\cal S}}=0\; .
\label{eqhdl}
\end{equation}
In the mean-field approximation, $\Gamma_2$
contains only the sunset-type diagram of Fig.\ 1a,
$ \Gamma_2[{\cal D}, {\cal S}] = - \frac{1}{4}
{\rm Tr} \left( {\cal D}\, \hat{\Gamma} \, {\cal S} \, \hat{\Gamma} \,
{\cal S} \right) $,
where $\hat{\Gamma}$ is the quark-gluon vertex.
The first equation (\ref{eqhdl}) gives rise 
to ${\cal D}^{-1}(K)= D^{-1}(K)+\Pi(K)$, with the self-energy 
$\Pi(K)$ given by Fig.\ 1b. The solid line 
represents the full quark propagator ${\cal S}\,$;
$K=(\vec k,\omega)$ is the Euclidean
four-momentum of the gluon and $\omega$ the discrete Matsubara frequency. 
At the stationary point (\ref{eqhdl}), 
the second term on the right-hand side of Eq.\ (\ref{eqtik}) cancels
the last term. We proceed by writing
\begin{subequations} \label{eqnick}
\begin{eqnarray}
{\cal S}(K)& =& {\cal S}_n(K)+\delta{\cal S}(K) \; , \\
{\cal D}^{-1}(K)& =& {\cal D}_n^{-1}(K)+\delta\Pi(K)\;,
\end{eqnarray}
\end{subequations}
where the subscript $n$ refers to quantities in the normal phase and 
$\delta{\cal S}(P)$ and $\delta\Pi(K)$ are functions of
the gap parameter $\Delta$ \cite{footnote}.
The CJT effective potential can be written as a sum of four terms
\begin{equation}
\Gamma=\Gamma_n+{\Gamma}_{cond}+{\Gamma}_{fluc}+\Gamma_{fluc}^\prime\;,
\label{eqerie}
\end{equation}
where
\begin{subequations}
\begin{eqnarray}
{\Gamma_{cond}}& =& \frac{k_B T}{2\, \Omega}
\left[ {\rm Tr} ({\cal D}_n{\delta}{\Pi})
-{\rm Tr}( S^{-1}\delta{\cal S}) \right. \nonumber \\
&   & \hspace*{0.8cm}
+ \left. {\rm Tr}\ln(1+{\cal S}_n^{-1}\delta{\cal S}) \right]\;, 
\label{eqbp}\\
{\Gamma}_{fluc}& =& \frac{k_B T}{2\,\Omega}
{\sum_{\vec k, \omega=0}} {\rm tr} \left\{
\ln \left[ 1+{\cal D}_n(K){\delta}{\Pi}(K)\right] \right. \nonumber \\
&  & \hspace*{2cm} - \left. {\cal D}_n(K){\delta}{\Pi}(K) \right\}\; ,
\label{eqmunif} \\
{\Gamma}_{fluc}^\prime & =& \frac{k_B T}{2\,\Omega}
{\sum_{\vec k, \omega\neq 0}} {\rm tr} \left\{
\ln\left[ 1+{\cal D}_n(K){\delta}{\Pi}(K) \right] \right. \nonumber \\
&  &  \hspace*{2cm} - \left. {\cal D}_n(K){\delta}{\Pi}(K) \right\}\;.
\label{eqmunifprime}
\end{eqnarray}
\end{subequations}
In our formulas Tr indicates summation over all indices including
momentum and energy while tr denotes summation
over all indices except momentum and energy.

The term $\Gamma_{cond}$ when expanded to the fourth power of 
$\Delta/(k_BT)$ gives rise to the Ginzburg-Landau
free energy \cite{lida,ioannis}. The entire quadratic term of 
$\Gamma$ is included in $\Gamma_{cond}$ and its coefficient vanishes 
at $T_c$ of Eq. (1). The term 
$\Gamma_{fluc}$ takes the explicit form
\begin{equation}
\label{fluct}
\Gamma_{fluc}=8\,k_BT\int\frac{d^3\vec k}{(2\pi)^3}\left\{
\ln\left[1+ \frac{m_A^2(k)}{k^2}\right]-\frac{m_A^2(k)}{k^2}\right\},
\end{equation}
where the factor 8 is the number of gluon colors. 
Furthermore, the small mixing with
the ordinary electromagnetic field is ignored
here. The momentum-dependent magnetic mass of the
gluons is given by $m_A^2(k) = f\left(\frac{k}{2\pi k_B
T}\right)/\lambda^2$,
where
\begin{equation}
\frac{1}{\lambda^2}=\frac{7\zeta(3)}{24\pi^4} \left( 
\frac{g\mu\Delta}{k_BT_c}\right)^2 \;.
\end{equation}
The function $f(y)$ becomes identical
to that of an electronic superconductor \cite{lif}
for $\Delta \ll k_BT$, i.e.,
\begin{equation}
f(y)=\frac{6}{7{\zeta}(3)}{\sum_{s=0}^{\infty}}{\int^{1}_0}dx
\frac{1-x^2}{(s+{\frac{1}{2}})[4(s+{\frac{1}{2}})^2+y^2x^2]}\;.
\label{eqkirkos}
\end{equation}
The limiting behavior of this function is $f(0)=1$ (London limit) and 
$f(y) \simeq 3{\pi}^3/[28{\zeta}(3)\,y]$ for $y \gg 1$ (Pippard limit).
The integrand of $\Gamma_{fluc}$ in the long-wavelength limit
becomes identical to the one used
in Ref.\ \cite{love}.

Because of the dynamical screening of the gluon propagator at 
nonzero Matsubara frequency, the term ${\Gamma}_{fluc}^\prime$ 
contributes terms of order higher than $O(g)$. Consequently,
it will be neglected in the following. The relevant free energy
density, expressed in terms of the gap energy $\Delta$ 
of the color-flavor locked (CFL) 
condensate and the temperature near $T_c$ reads
\begin{eqnarray}
\lefteqn{ \Gamma-\Gamma_n  =  \frac{6\mu^2}{\pi^2}\,t \, \Delta^2
+\frac{21\zeta(3)}{4\pi^4}\left(\frac{\mu}{k_BT_c}\right)^2\Delta^4
}  \nonumber \\
& +& 
32\pi(k_BT_c)^4\, F\left(\frac{1}{4\pi^2k_B^2T_c^2\lambda^2}\right)
\equiv \gamma(t, \Delta) \; ,
\label{free}
\end{eqnarray}
where $t\equiv (T-T_c)/T_c$, and
\begin{equation}
F(z)={\int^{\infty}_0}dx\, x^2 \left\{ \ln \left[1+{\frac{z}{x^2}}f(x)\right]
-\frac{z}{x^2}f(x) \right\}\;.
\label{eqrigas}
\end{equation}
The asymptotic behavior of the function $F(z)$ can be inferred from
that of $f(z)$,
\begin{equation}
F(z)\simeq \left\{ \begin{array}{ll} 
 - \frac{\pi}{3}z^{3/2}& {\rm for}\; z \ll 1 \;, \\
 - \frac{{\pi}^3}{28{\zeta}(3)}z
 \left[\ln\left(\frac{3\pi^3}{28 {\zeta}(3)}z \right)+{\rm const}\right]
 & {\rm for}\; z\gg 1\; . \end{array} \right.
\label{asymptF}
\end{equation}
The $z \ll 1$ behavior of the function, when
substituted into Eq. (\ref{free}), gives rise to 
the well-known $\Delta^3$ term of Ref.\ \cite{ma}.
The transition temperature $T_c^*$ and the
value of the gap $\Delta$ at the first-order phase transition
are determined from the nontrivial solution of the pair of equations
\begin{equation}
\gamma(t^*,\Delta) \equiv 0\quad, \qquad 
\frac{\partial \gamma (t^*, \Delta)}{\partial \Delta^2} \equiv 0.
\label{eqwarp}
\end{equation}
Because of the trivial dependence of $\gamma(t^*,\Delta)$ on $t^*$, 
we can combine the two equations into one
\begin{equation}
{\cal F}\left(\frac{1}{4\pi^2k_B^2T_c^2\lambda^2}\right)=
\frac{216{\pi}^7}{7{\zeta}(3)g^4}\left(\frac{k_B T_c}{\mu}\right)^2\;,
\label{eqvrema}
\end{equation}
with ${\cal F}(z)=-F^{\prime}(z)/z+F(z)/z^2$. It follows
from the explicit form of $F(z)$ that the
function ${\cal F}(z)$ is monotonically
decreasing for $z>0$, i.e.
\begin{equation}
{\cal F}^\prime(z)=-\frac{2}{z^3}\int_0^\infty dx\,x^2\left[\ln\frac{1}{1-w}
-w-\frac{1}{2}w^2\right]\leq 0\;,
\end{equation}
where $w=zf(x)/[x^2+zf(x)]$.
This, together with the asymptotic behavior
\begin{equation}
{\cal F}(z)\simeq \left\{ \begin{array}{ll}
\frac{\pi}{6}\,z^{-1/2} & {\rm for}\;  z \ll 1\; , \\
\frac{\pi^3}{28{\zeta}(3)}\,z^{-1} & {\rm for}\; z \gg 1\, ,
\end{array} \right.
\label{asymptCF}
\end{equation}
implied by Eq.\ (\ref{asymptF}), demonstrates the existence and the uniqueness
of the solution to Eq.\ (\ref{eqvrema}) and therefore to the set of Eqs.\ 
(\ref{eqwarp}). Furthermore, we can show that $t^*>0$. In the
weak-coupling limit, the right-hand side of Eq.\ (\ref{eqvrema}) becomes
small and the
solution lies on the side of a large argument of ${\cal F}(z)$. It follows 
from Eq.\ (\ref{asymptCF}) that
\begin{equation}
\left(\frac{\Delta}{k_BT_c}\right)^2=\frac{\pi^2}{63\zeta(3)}g^2
\label{eqgap}
\end{equation}
at the transition, which implies that $\lambda \ll \xi_0$.
Substituting Eq.\ (\ref{eqgap}) into either one of Eqs. (\ref{eqwarp}), 
we obtain 
\begin{equation}
\label{eqfirst1}
t^*\equiv\frac{T_c^*-T_c}{T_c}\simeq\frac{g^2}{72}\Big[\ln\left(
\frac{\mu}{k_BT_c}\right)^2+{\rm const.}\Big]\,,
\end{equation}
which, upon substitution of Eq.\ (\ref{pairing}), yields 
Eq.\ (\ref{eqfirst0}) to leading order in $g$.

The strength of the first-order phase transition measured by Eq.\
(\ref{eqfirst1})
or Eq.\ (\ref{eqfirst0}), though robust and vanishing in the limit $g\to
0$, is much stronger than that estimated in Ref.\ \cite{baym}. 
The extrapolation of this formula to the accessible baryon 
density inside a neutron star, say $\mu=500$ MeV, gives rise to 
$(T_c^*-T_c)/T_c\simeq 1.8$ for $g=3.1$, given by the one-loop formula with 
$\Lambda_{\rm QCD}=200$ MeV. This estimate, though 
inconsistent with the assumption $t \ll 1$, indicates that the gauge field 
fluctuations cannot be neglected for the color-superconducting 
phase transition at moderate baryon density.     

It is well known in solid state physics that the
local coupling to the electromagnetic 
gauge potential in the Ginzburg-Landau free energy functional 
cannot be sustained far away
from $T_c$ and the condition for locality could be 
more stringent than that of
the Ginzburg-Landau expansion of the condensation
energy, i.e., $\Delta/(k_BT_c)\ll 1$, for a strong type-I 
superconductor \cite{gorkov}. What we found above
demonstrates this subtlety.
It is interesting to examine the local approximation employed
in Ref. \cite{ma} for the electronic 
superconductor using the
formalism developed here. The condensation energy and the 
fluctuation energy take the form
\begin{eqnarray}
\gamma(t,\Delta)& =& \frac{k_F^2}{2\pi^2v_F}t\Delta^2
+\frac{7\zeta(3)k_F^2}{32\pi^4v_Fk_B^2T_c^2}\Delta^4 \nonumber \\
&   & +\frac {4\pi(k_BT_c)^4}{v_F^3}
F\left(\frac{v_F^2}{4\pi^2k_B^2T_c^2\lambda^2}\right)\;,
\label{free1}
\end{eqnarray}
with $\lambda^{-2}=\frac{7\zeta(3)v_F}{12\pi^4}[ek_F\Delta/
(k_BT_c)]^2$ with $k_F$ and $v_F$ the Fermi momentum and the Fermi 
velocity. The zero-temperature coherence length 
$\xi_0\sim v_F/(k_BT_c)$. 
Correspondingly, Eq.\ (\ref{eqvrema}) is replaced by
\begin{equation}
{\cal F}\left(\frac{v_F^2}{4\pi^2k_B^2T_c^2\lambda^2}\right)
=\frac{\pi^2\kappa^2}{16{\alpha_e}v_F}\;,
\label{eqfirst_em}
\end{equation}
where $\alpha_e\simeq \frac{1}{137}$ is the fine structure 
constant and 
\begin{equation}
\kappa=3\sqrt{\frac{2}{7\zeta(3)\alpha_e}}
\Big(\frac{\pi}{v_F}\Big)^{\frac{3}{2}}
\frac{k_BT_c}{k_F}
\end{equation} 
is the Ginzburg-Landau parameter. 
The validity of the local coupling approximation relies on a large value 
of the right-hand side of Eq.\ (\ref{eqfirst_em}), which
implies that 
\begin{equation}
\kappa \gg \frac{4\sqrt{\alpha_ev_F}}{\pi}.
\label{eqsec}
\end{equation}
For typical metals  $v_F\sim\alpha_e$ and consequently
Eq.\ (\ref{eqsec}) implies that $\kappa \gg 0.009$. The 
local approximation, though marginal for the strongest
type-I material like aluminium ( $\kappa=0.01\sim 0.02$ ), works practically 
for all laboratory-prepared type-I materials.  

In this letter, we have incorporated consistently the fluctuations of the
gauge field into the free energy 
of a homogeneous CFL color superconductor in mean-field approximation.
We determined the temperature of the fluctuation-induced
first-order phase transition to the color-superconducting phase
in weak coupling. We find that the typical momentum of the 
fluctuations corresponds to the Pippard limit of the magnetic self-energy of 
gluons and the conventional local coupling
approximation of the fluctuation, though applicable for the metallic 
superconductors, breaks down for color superconductivity.

The phase transition to the CFL phase is accompanied by a spontaneous 
breaking of the chiral symmetry, which is of first order 
because of instantons \cite{PW}.
What we found in this letter is that the fluctuations of the 
gauge fields induce a much stronger first-order phase transition
in weak coupling and that the mechanism is not limited to three flavors.

\section*{Acknowledgments}
H.-c.R.\ is indebted to D.T.\ Son for
instructive discussions. H.-c.R.\ and D.H.R.\ thank
the I.N.T. of University of Washington for 
hospitality where 
part of this work was done. We would like to thank K.\ Iida for an interesting 
correspondence. We are grateful to B. Halperin 
for pointing out a mistake on the discussion of the metallic superconductor 
in the original manuscript. The work of I.G.\ and H.-c.R.\
is supported in 
part by the US Department of Energy under grants
DE-FG02-91ER40651-TASKB 
and the work of D.H.\ is supported in part by 
Alexander von Humboldt-Foundation and by the 
National Natural Science Foundation of China under grants 1000502 and 10135030.

\begin{figure}[t]
\includegraphics[width=8cm]{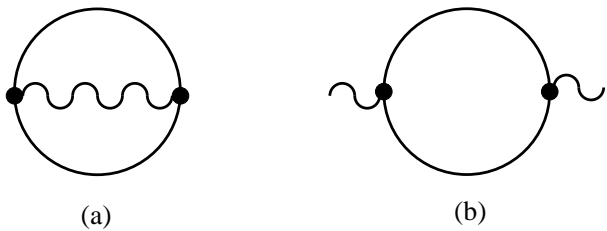}
\caption{(a) The mean field approximation to $\Gamma_2$.
(b) The gluon self-energy.}
\label{fig1}
\end{figure}


\begin{thebibliography}{99}
\bibitem{love}
For early works, see: D.\ Bailin and A.\ Love, Phys.\ Rept.\ 107 (1984) 325,
and references therein.
\bibitem{reviews} K.\ Rajagopal and F.\ Wilczek, hep-ph/0011333,
to appear in B.L.\ Ioffe Festschrift, {\it At the Frontier of Particle
Physics/Handbook of QCD}, M.\ Shifman ed., (World Scientific 2001);
M.\ Alford, Ann.\ Rev.\ Nucl.\ Part.\ Sci.\ 51 (2001) 131;
T.\ Sch\"afer, hep-ph/0304281; 
H.-c.\ Ren, hep-ph/0404074, and references therein.
\bibitem{review}
D.H.\ Rischke, Prog.\ Part.\ Nucl.\ Phys.\ 52 (2004) 197.
\bibitem{NJLreview} For a review on the NJL-model approach to
color superconductivity, see: M.\ Buballa, hep-ph/0402234.
\bibitem{arw}
M.\ Alford, K.\ Rajagopal, and F.\ Wilczek, Nucl.\ Phys.\ B 537 (1999) 443;
R.\ Rapp, T.\ Sch\"afer, E.V.\ Shuryak, and M.\ Velkovsky,
Phys.\ Rev.\ Lett.\ 81 (1998) 53.
\bibitem{son}
D.T.\ Son, Phys.\ Rev.\ D 59 (1999) 094019.
\bibitem{schafer}
T.\ Sch\"afer and F.\ Wilczek, Phys.\ Rev.\ D 60 (1999) 114033.
\bibitem{rischke}
R.D.\ Pisarski and D.H.\ Rischke, Phys.\ Rev.\ D 61 (2000) 051501.
\bibitem{hong}
D.K.\ Hong, V.A.\ Miransky, I.A.\ Shovkovy, and L.C.R.\ Wijewardhana,
Phys.\ Rev.\ D 61 (2000) 056001,
[Erratum-ibid.\ D 62 (2000) 059903].
\bibitem{brown}
W.E.\ Brown, J.T.\ Liu, and H.-c.\ Ren, Phys.\ Rev.\ D 61 (2000) 114012, 
Phys.\ Rev.\ D 62 (2000) 054016, 054013.
\bibitem{wang}
Q.\ Wang and D.H.\ Rischke, Phys.\ Rev.\ D 65 (2002) 054005.
\bibitem{lida}
K.\ Iida and G.\ Baym, Phys.\ Rev.\ D 63 (2001) 074018.
\bibitem{ioannis}
I.\ Giannakis and H.-c.\ Ren, Phys.\ Rev.\ D 65 (2002) 054017.
\bibitem{ma}
B.I.\ Halperin, T.C.\ Lubensky, and S.\ Ma, Phys.\ Rev.\ Lett.\ 32 (1974) 292.
\bibitem{gia}
I.\ Giannakis and H.-c.\ Ren, Nucl.\ Phys.\ B 669 (2003) 462.
\bibitem{baym}
T.\ Matsuura, K.\ Iida, T.\ Hatsuda, and G.\ Baym, Phys.\ Rev. \ D 69
(2004) 074012.
\bibitem{cjt}
J.M.\ Cornwall, R.\ Jackiw, and E.\ Tomboulis, Phys.\ Rev.\ D 10 (1974) 2428.
\bibitem{footnote}
In terms of the parameter $\Delta$ of this letter, the energy gap of the 
fermionic quasiparticle excitations is $\Delta$ 
(8-fold) and $2\Delta$ (1-fold). See for example, D.H.\ Rischke, Phys.\ Rev.\ 
D 62 (2000) 054017.
\bibitem{lif}
E.M.\ Lifshitz and L.P.\ Pitaevskii, {\it Statistical Physics}
(Pergamon Press).
\bibitem{gorkov}
L.P. Gorkov, Soviet Physics JETP 36(9), No. 6, (1959) 1364.
\bibitem{PW}
R.\ Pisarski and F.\ Wilczek, Phys.\ Rev.\ D 29 (1984) 338.
\end{thebibliography}
\end{document}